\baselineskip 16pt
\font\carbig=cmr10 scaled 2074

\magnification=\magstep0
\hsize=15.0truecm
\hoffset=0.6truecm
\vsize=22.0truecm
\hyphenation{IL-LU-MI-NA-TI}

\def\sqr#1#2{{\vcenter{\vbox{\hrule height.#2pt
    \hbox{\vrule width.#2pt height#1pt \kern#1pt
    \vrule width.#2pt}\hrule height.#2pt}}}}

\def\boxit#1{\vbox{\hrule\hbox{\vrule\kern5pt
     \vbox{\kern5pt#1\kern3pt}\kern5pt\vrule}\hrule}}

\def\E{{\bf E}}
\def\e{{\rm e}}

\def\RE{{\bf R}}

\pageno=1
\item{}
\vskip 1cm
\centerline{\carbig Controlled quantum evolutions}
\centerline{\carbig and transitions}
\vskip 12pt
\centerline{\bf Nicola Cufaro Petroni}
\par
\centerline{{\it INFN Sezione di Bari, INFM Unit\`a di Bari and }}
\par
\centerline{{\it Dipartimento Interateneo di Fisica
                 dell'Universit\`a e del Politecnico di Bari,}}
\par
\centerline{{\it via Amendola 173, 70126 Bari (Italy)}}
\par
\centerline{CUFARO@BA.INFN.IT}
\vskip 7pt
\centerline{{\bf
Salvatore De Martino, Silvio De Siena and Fabrizio Illuminati}}
\par
\centerline{{\it INFM Unit\`a di Salerno,}}
\par
\centerline{{\it INFN Sezione di Napoli -
Gruppo collegato di Salerno and}}
\par
\centerline{{\it Dipartimento di Fisica dell'Universit\`a di Salerno}}
\par
\centerline{{\it via S.Allende, 84081 Baronissi, Salerno (Italy)}}
\par
\centerline{DEMARTINO@PHYSICS.UNISA.IT, DESIENA@PHYSICS.UNISA.IT}
\par
\centerline{ILLUMINATI@PHYSICS.UNISA.IT} \vskip 1.3cm \noindent
{\bf ABSTRACT: {\it We study the non stationary solutions of
Fokker--Planck equations associated to either stationary or non
stationary quantum states. In particular we discuss the stationary
states of quantum systems with singular velocity fields. We
introduce a technique that allows arbitrary evolutions ruled by
these equations to account for controlled quantum transitions. As
a first signficant application we present a detailed treatment of
the transition probabilities and of the controlling
time--dependent potentials associated to the transitions between
the stationary, the coherent, and the squeezed states of the
harmonic oscillator.}}

\vskip 1.3cm \noindent {\bf 1. Introduction} \vskip 10pt \noindent
In a few recent papers [1] the analogy between diffusive classical
systems and quantum systems has been reconsidered from the
standpoint of the stochastic simulation of quantum mechanics [2],
[3], [4] and particular attention has been devoted there to the
evolution of the classical systems associated to a quantum wave
function when the conditions imposed by the stochastic variational
principle are not satisfied (non extremal processes). The problem
studied in those papers was the convergence of an arbitrary
evolving probability distribution, solution of the Fokker--Planck
equation, toward a suitable quantum distribution. In [1] it was
pointed out that, while the correct convergence is achieved for a
few quantum examples, these results cannot be considered general
as shown in some counterexamples: in fact not only for particular
non stationary wave functions (as for a minimal uncertainty
packet), but also for stationary states with nodes one does not
recover in a straightforward way the correct quantum asymptotic
behaviour. For stationary states with nodes the problem is that
the corresponding velocity field to consider in the Fokker--Planck
equation shows singularities at the locations of the nodes of the
wave function. These singularities effectively separate the
available interval of the configurational variables into non
communicating sectors which trap any amount of probability
initially attributed and make the system non ergodic.
\par
In a more recent paper [5] it has been shown that
for transitive systems with stationary velocity
fields (as, for example, a stationary state without nodes)
we always have an exponential convergence to
the correct quantum probability distribution
associated to the extremal process,
even if we initially start from an
arbitrary non extremal process.
These results can also be extended to
an arbitrary stationary state if we
separately consider the process as
confined in every region of the configuration space
between two subsequent nodes.

In the same paper [5] it has been further remarked that while the
non extremal processes should be considered virtual, as the non
extremal trajectories of classical Lagrangian mechanics, they can
however become physical, real solutions if we suitably modify the
potential in the Schr\"odinger equation. The interest of this
remark lies not so much in the fact that non extremal processes
are exactly what is lacking in quantum mechanics in order to
interpret it as a totally classical theory of stochastic processes
(for example in order to have a classical picture of a double slit
experiment [6]), but rather in the much more interesting
possibility of engineering and controlling physically realizable
evolutions of quantum states. This observation would be of great
relevance, for instance to the study and the description of (a)
transitions between quantum states (b) possible models for quantum
measurements [3] and (c) control of the dynamics of quantum--like
systems (for instance charged beams in particle accelerators) [7].

In particular, case (c) is being studied in the framework of Nelson
stochastic mechanics which is an independent and self--consistent
reformulation of quantum mechanics [2], [3] and can be applied in
other areas of physical phenomenology. For instance it can
usefully account for systems not completely described by the
quantum formalism, but whose evolution is however strongly
influenced by quantum fluctuations, i.e. the so--called mesoscopic
or quantum--like systems. This behaviour characterizes, for
example, the beam dynamics in particle accelerators and there is
evidence that it can be described by the stochastic formalism of
Nelson diffusions [1], [7] since in these quantum--like systems,
trajectories and transition probabilities acquire a clear physical
meaning, at variance with the case of quantum mechanics.

On the other hand, quantum behaviours can be simulated
by means of classical stochastic processes
in a by now well defined and established framework [2].
A stochastic variational principle provides a
foundation for that, in close analogy with
classical mechanics and field theory [3]. In this scheme the
deterministic trajectories of classical mechanics are replaced
by the random trajectories of diffusion processes in
configuration space. The programming
equations derived from the stochastic
variational principle
are formally identical to the
equations of the Madelung fluid [8], the hydrodynamical equivalent
of Schr\"odinger equation [9].
On this basis, it is possible
to develop a model whose phenomenological predictions coincide
with those of quantum mechanics for all the experimentally
measurable quantities. Within this interpretative code
stochastic mechanics is nothing but a quantization procedure,
different from the canonical one only formally, but completely
equivalent from the point of view of the physical consequences: a
probabilistic simulation of quantum mechanics, providing a bridge
between this fundamental physical theory and stochastic
differential calculus.
However, it is well known that the central
objects in the theory of classical stochastic processes,
namely the transition probability densities,
seldom play any observable role in stochastic mechanics
and must be considered as a sort of gauge variables.
Several generalizations of Nelson stochastic quantization
have been recently proposed to allow for the observability
of the transition probabilities: for instance,
stochastic mechanics could be modified by means of non
constant diffusion coefficients [1]; alternatively, it has been
suggested that the stochastic evolution might be modified
during the measurement process [10].

The aim of the present paper is instead to show how
the transition probabilities associated to Nelson diffusion
processes can play a very useful role in standard quantum
mechanics, in particular with regard to describing and
engineering the dynamics of suitably controlled quantum
evolutions and transitions. More precisely, we consider
the following problem in the theory of quantum control:
{\it given an initial probability
distribution $\rho_i$
associated to an arbitrarily assigned
quantum state $\psi_i$,
we study its time evolution with the drift
associated to another arbitrarily assigned
quantum state $\psi_f$,
to determine the controlling time--dependent potential
$V_{c}(x,t)$ such that, I) at any instant of time the evolving
probability distribution is that associated to the wave
function solution of the Schr\"odinger equation in the potential
$V_{c}(x,t)$, and that, II) asymptotically in
time the evolving distribution converges to the distribution
$\rho_f$ associated to $\psi_f$}.

After introducing the formalism of Nelson stochastic mechanics to
describe quantum evolutions in Sections 2 and 3, we provide in
Sections 3 and 4 a self--contained review of the Sturm--Liouville
problem for the Fokker--Planck equation and the techniques of
solution for the Nelson diffusions associated both to
nonstationary and stationary quantum states. In Section 5 we
discuss in detail the example of the harmonic oscillator,
explicitely solving for the transition probability densities of
the ground and of the low lying excited states. Sections 6, 7 and
8 are devoted to the study and the solution of the problem
outlined above, discussing the potentials associated to the
definition of controlled quantum evolution, and modelling
transitions. Two explicit examples are studied in detail: the
controlled transition between the invariant probability densities
associated to the ground and the first excited state of the
harmonic oscillator, and the controlled evolution between pairs of
coherent or squeezed wave packets. In these cases the problem can
be solved completely, yielding the explicit analytic form of the
evolving transition probabilities and of the evolving controlling
potentials at all times. Finally, in Section 9 we present our
conclusions and discuss possible future extensions and
applications of the technique introduced in the present paper,
with regard to the discussion of anharmonic quantum and
quantum--like systems, the role of instabilities in the initial
conditions, and the implementation of optimization procedures.

\vskip 30pt
\noindent
{\bf 2. Fokker-Planck equations and quantum systems}
\vskip 10pt
\noindent
Here we will recall a few notions of stochastic mechanics in order
to fix the notation. The configuration of a classical particle is
promoted to a vector Markov process $\xi(t)$
taking values in $\RE^3$.
This process is characterized by a probability
density $\rho ({\bf r},t)$ and a transition probability density
$p({\bf r},t|\,{\bf r'},t')$ and its components
satisfy an It\^o stochastic
differential equation of the form
$$d\xi_j(t)=v_{(+)j}\bigl(\xi(t),t\bigr)dt+d\eta_j(t)
\, , \eqno(2.1)$$
where $v_{(+)j}$ are the components of the forward velocity field.
However here the fields $v_{(+)j}$ are not given a priori, but
play the role of dynamical variables and are consequently
determined by imposing a specific dynamics.
The noise $\eta(t)$ is a standard Wiener
process independent of $\xi(t)$ and such that
$$\E_t\bigl(d\eta_j(t)\bigr)=0\,,\qquad
  \E_t\bigl(d\eta_j(t)\,d\eta_k(t)\bigr)=2 \,
D \, \delta_{jk}\, dt \, , \eqno{(2.2)}$$
where $d\eta(t)=\eta(t+dt)-\eta(t)$ (for $dt>0$), $D$ is the
diffusion coefficient,
and $\E_t$ are the conditional expectations
with respect to $\xi(t)$.
In what follows, for sake of notational simplicity,
we will limit ourselves
to the case of one dimensional trajectories, but the
results that will be obtained can be immediately
generalized to any number of dimensions.
We will suppose for the time being that the forces
will be defined by means of purely
configurational potentials,
possibly time--dependent $V(x,t)$. A
suitable definition of the Lagrangian
and of the stochastic action
functional for the system described
by the dynamical variables
$\rho$ and $v_{(+)}$ allows to select,
the processes which
reproduce the correct quantum
dynamics [2], [3]. In fact, while
the probability density $\rho (x,t)$
satisfies, as
usual, the forward Fokker--Planck
equation associated to the
stochastic differential equation (2.1) $$\partial_t \rho = D
\partial_x^2 \rho -\partial_x ( v_{(+)} \rho ) = \partial_x ( D
\partial_x \rho - v_{(+)} \rho ) \, ,
                                       \eqno{(2.3)}$$
the following choice for the Lagrangian field
$$L(x,t)={m\over2}v_{(+)}^2 (x,t) + m D \partial_x v_{(+)}(x,t) -
V(x,t) \, , \eqno{(2.4)}$$
enables to define a stochastic action functional
$${\cal A}=\int_{t_0}^{t_1}\E \big[L\big(\xi(t),t\big)
\big] \, dt \, , \eqno{(2.5)}$$
which leads, through the stationarity condition
$\delta {\cal A}=0$,
to the equation
$$\partial_tS +
{(\partial_xS)^2\over2m} + V \, - 2mD^2 \,
   {\partial^2_x\sqrt{\rho } \over \sqrt{\rho }} = 0 \, .
\eqno{(2.6)}$$
The field $S(x,t)$ is defined as
$$S(x,t)=-\int_t^{t_1}
  \E\left[ L\big(\xi(s),s\big)\,\big|\,\xi(t)=x \right] \, ds
  +\E\left[ S_1\big(\xi(t_1)\big)\,\big|\,\xi(t)=x \right]
\, , \eqno{(2.7)}$$
where
$S_1(\,\cdot\,)=S(\,\cdot\,,t_1)$
is an arbitrary final condition.
By introducing the function
$R(x,t) \equiv \sqrt{\rho(x,t)}$
and the de Broglie Ansatz
$$\psi(x,t)=R(x,t)\,{\rm e}^{iS(x,t)/2mD} \, \, ,
\eqno{(2.8)}$$
equation (2.6) takes the form
$$\partial_tS+{(\partial_xS)^2\over2m}+V\,-\,{2mD^{2}}\,
        {\partial^2_xR\over R}=0 \, , \eqno{(2.9)}$$
and the complex function $\psi$
satisfies the Schr\"odinger--like
equation
$$i(2mD)\partial_t\psi=\hat H\psi=
-\,{2mD^{2}}\,\partial_x^2\psi
   + V\psi \, . \eqno{(2.10)}$$
If the diffusion coefficient is chosen to be $$D={\hbar\over 2m}\,
, \eqno{(2.11)}$$ we recover exactly the Schr\"odinger equation of
quantum mechanics. Different choices of $D$ allow instead to
describe the effective quantum--like dynamics of more general
systems.
\par
On the other hand,
if we start from the (one--dimensional) Schr\"odinger equation
(2.10) with the de Broglie Ansatz (2.8) and the diffusion
coefficient (2.11), separating the real and the imaginary parts as
usual in the hydrodynamical formulation [8], we recover equations
(2.3) and (2.6) with $\rho = R^2 =|\psi|^2$ and the forward
velocity field $$v_{(+)}(x,t) = {1\over m} \,
\partial_xS+{\hbar\over 2m}\,\partial_x(\ln R^2) \, .
\eqno{(2.12)}$$

\vskip 30pt
\noindent
{\bf 3. The Sturm--Liouville problem and the
solutions of the Fokker--Planck equation}
\vskip 10pt
\noindent
Let us recall here (see for example [11]) a few
generalities about the techniques of solution of the
Fokker--Planck equation with $D$ and $v_{(+)}$ two
time--independent continuous and differentiable
functions defined for
$x\in[a,b]$ and $t\geq t_0$, such that
$D(x)>0$, and $v_{(+)}(x)$ has no
singularities in $(a,b)$.
The Fokker--Planck equation then reads
$$\partial_t \rho =\partial_x^2(D\rho )-\partial_x(v_{(+)}
\rho )=
\partial_x\bigl[\partial_x(D\rho )-v_{(+)}
\rho \bigr] \, . \eqno{(3.1)}$$
The conditions imposed on the
probabilistic solutions are of course
$$\eqalign{\rho (x,t)\geq0\,,&\qquad\qquad
a<x<b\,,\;t_0\leq t\,,\cr
\int_a^b \rho(x,t)\,dx=1\,,&\qquad\qquad t_0\leq t\,,\cr}
\eqno{(3.2)}$$
and from the form of (3.1)
the second condition takes the form
$$\bigl[\partial_x(D\rho )-v_{(+)} \rho
\bigr]_{a,b}=0\,,\qquad t_0\leq t\,.\eqno{(3.3)}$$
Suitable initial conditions
will be added to produce the required
evolution: for example the
transition probability density
$p(x,t|x_0,t_0)$ will be
selected by the initial condition
$$\lim_{t\to t_0^+}\rho (x,t)= \rho_{in}
(x)=\delta(x-x_0)\,.\eqno{(3.4)}$$
It is also possible to show by
direct calculation that
$$h(x)=N^{-1}\,\e^{-\int[D'(x)
-v_{(+)}(x)]/D(x)\,dx} \, \, , \qquad
\left(  N=\int_a^b\e^{-\int[D'(x)-v_{(+)}(x)]/D(x)\,dx} \, dx
\right) \eqno{(3.5)}$$
is always an invariant
(time independent) solution of (3.1)
satisfying the conditions (3.2)
(here the prime symbol denotes differentiation).
One should observe however that
relation (3.1) is not in the
standard self--adjoint form [12];
this fact notwithstanding,
if we define the new function $g(x,t)$ by means of
$$\rho (x,t)=\sqrt{h(x)}\,g(x,t) \, , \eqno{(3.6)}$$
it is easy to show
that $g(x,t)$ obeys an equation of the form
$$\partial_tg={\cal L}g \, , \eqno{(3.7)}$$
where the operator ${\cal L}$ acting on positive
normalizable functions $\varphi(x)$ and defined by
$${\cal L}\varphi=
     {{\rm d}\over {\rm d}x}\left[r(x)\,
{{\rm d}\varphi(x)\over {\rm d}x}\right]-q(x)\varphi(x)\,,
               \eqno{(3.8)}$$
with
$$\eqalign{r(x)&=D(x)>0\,,\cr
    q(x)&={\bigl[D'(x)-v_{(+)}(x)\bigr]^2\over4D(x)}\,-\,
{\bigl[D'(x)-v_{(+)}(x)\bigr]'\over2}\,,\cr}\eqno{(3.9)}$$
is now self--adjoint.
By separating the variables by means of
$g(x,t)=\gamma(t)G(x)$
we have $\gamma(t)={\rm e}^{-\lambda t}$ while $G$ must
be solution of a typical
Sturm-Liouville problem associated to
the equation
$${\cal L}G(x)+\lambda G(x)=0 \, , \eqno{(3.10)}$$
with the boundary conditions
$$\eqalign{&\bigl[D'(a)-v_{(+)}(a)\bigr]G(a)+2D(a)G'(a)=0\,,\cr
           &\bigl[D'(b)-v_{(+)}(b)\bigr]G(b)+2D(b)G'(b)=0\,.\cr}
     \eqno{(3.11)}$$
It is easy to see
that $\lambda=0$ is always an eigenvalue for the
problem (3.10) with (3.11), and that the corresponding
eigenfunction is $\sqrt{h(x)}$ as defined from (3.5).
\par
For the differential problem (3.10)
with (3.11) we have that [12]
the simple eigenvalues $\lambda_n$
will constitute an infinite,
monotonically increasing sequence and the corresponding
eigenfunction $G_n(x)$ will have $n$ simple
zeros in $(a,b)$. This means
that $\lambda_0=0$, corresponding
to the eigenfunction $G_0(x)=\sqrt{h(x)}$
which never vanishes in $(a,b)$, is the
lowest eigenvalue, and that all the other
eigenvalues are strictly positive.
Moreover the eigenfunctions will
form a complete orthonormal set
of functions in $L^2\bigl([a,b]\bigr)$
[13]. As a consequence, the general
solution of equation (3.1)
satisfying the conditions (3.2) will have the form
$$\rho (x,t)=\sum_{n=0}^{\infty}c_n\e^{-\lambda_nt}
\sqrt{h(x)}G_n(x) \, , \eqno{(3.12)}$$
with $c_0=1$ for normalization
(remember that $\lambda_0=0$). The
coefficients $c_n$ for a
particular solution are selected by an
initial condition $$\rho (x,t_0^+) =
\rho_{in}(x) \, , \eqno{(3.13)}$$
and are then calculated from
the orthonormality relations as
$$c_n=\int_a^b\rho_{in}(x) \,
{G_n(x)\over\sqrt{h(x)}}\,dx\,.\eqno{(3.14)}$$
In particular for the transition
probability density we have from
(3.4) that
$$c_n={G_n(x_0)\over\sqrt{h(x_0)}}\,.\eqno{(3.15)}$$
\par\noindent
Since $\lambda_0=0$
and $\lambda_n>0$ for $n>1$, the general
solution (3.12) of (3.1) has a
precise time evolution: all the
exponential factors in (3.12)
vanish as $t\to+\infty$ with the only
exception of the term $n=0$ which
is constant, so that
exponentially fast we will
always have $$\lim_{t\to+\infty}\rho
(x,t)=c_0\sqrt{h(x)}G_0(x)=h(x)\, . \eqno{(3.16)}$$
Therefore the
general solution will always
relax in time toward the invariant
solution $h(x)$.
As a consequence the eigenvalues $\lambda_{n}$
which solve the Sturm--Liouville
problem define the physical time
scales of the decay. By the structure
of equations (3.7)--(3.11)
we see that tuning the choice
of the physical parameters that
enter in the diffusion coefficient
and in the forward velocity
field allows for
different sets of eigenvalues which define
different sets of time scales.
Hence, the rate of convergence can
be fixed as to yield fast decay,
slow decay, or even, on proper
observational scales,
quasi metastable behaviours,
according to
what kind of physical evolution
between quantum states one wants
to realize.
This point will be further
discussed and elucidated in Section 6.

\vskip 30pt
\noindent
{\bf 4. Processes associated to stationary quantum states}
\vskip 10pt
\noindent
Let us consider
now a Schr\"odinger equation (2.10)
with a time--independent
potential $V(x)$ which gives rise
to a purely discrete spectrum
and bound, normalizable states.
Let us introduce the following
notations for stationary states,
eigenvalues and eigenfunctions:
$$\eqalign{\psi_n(x,t)&=\phi_n(x)\,
{\rm e}^{-iE_nt/\hbar} \, , \cr
      \hat H\phi_n&=-{\hbar^2\over 2m}\,\phi''_n+V\phi_n =
E_n\phi_n \, . \cr }
\eqno{(4.1)}$$
Taking into account
relation (2.11), the previous eigenvalue
equation can also be recast in the
following form
$$D\phi''_n={V-E_n\over\hbar}\,\phi_n\,.\eqno{(4.2)}$$
For these
stationary states the probability
densities are the time--independent,
real functions
$$\rho_n(x)=|\psi_n(x,t)|^2=\phi_n^2(x)\,,
\eqno{(4.3)}$$
while the phase and the amplitude of $\psi_n$ are
from (2.8)
$$S_n(x,t)=-E_nt\,,\qquad R_n(x,t)=\phi_n(x)\,,
\eqno{(4.4)}$$ so that
the associated velocity fields are from (2.12)
$$v_{(+)n}(x)=2D\,{\phi'_n(x)\over\phi_n(x)}\,.
\eqno{(4.5)}$$
Each $v_{(+)n}$ is time--independent and presents
singularities in the nodes of the associated
eigenfunction.
Since the
$n$--th eigenfunction of a quantum
system with bound states has
exactly $n$ simple nodes
$x_1,\dots,x_n$,
the coefficients of the Fokker-Planck equation
(2.3) are not defined in these $n$ points
and it is necessary to
solve it in separate intervals
by imposing the correct boundary
conditions connecting the different sectors.
In fact these
singularities effectively
separate the real axis in $n+1$
sub--intervals with walls
impenetrable to the probability current.
Hence the process will not
have an unique invariant measure and
will never cross the boundaries
fixed by the singularities of
$v_{(+)}(x)$: if the process starts in one
of the sub--intervals, it
will always remain there [14].
\par
As a consequence, the normalization integral (3.2)
(with $a=-\infty$ and $b=+\infty$) is
the sum of $n+1$ integrals
over the sub--intervals $[x_k,x_{k+1}]$
with $k=0,1,\dots,n$ (where
we understand, to unify the notation,
that $x_0=-\infty$ and
$x_{n+1}=+\infty$).
Hence for $n\geq1$ equation (2.3) must be
restricted in
each interval $[x_k,x_{k+1}]$
with the integrals
$$\int_{x_k}^{x_{k+1}}\rho (x,t)\,dx
\, , \eqno{(4.6)}$$
constrained to
a constant value for $t\geq t_0$.
This constant is not, in
general, equal to one
(only the sum of these $n+1$ integrals
amounts to one) and,
since the separate intervals cannot
communicate, it will be fixed by
the choice of the initial
conditions.
Therefore, due to the
singularities appearing in the forward velocity
fields $v_{(+)n}$ for $n\geq 1$,
we deal with a Fokker--Planck problem
with barriers.
The boundary conditions associated to (2.3) then
require the conservation of probability
in each sub--interval $[x_k,x_{k+1}]$,
i.e.
the vanishing of the probability
current at the end points of the
interval: $$\bigl[D\partial_x \rho- v_{(+)} \rho
\bigr]_{x_k,x_{k+1}}
= 0\,,\qquad t\geq t_0\,.  \eqno{(4.7)}$$
To obtain a particular solution
one must specify the initial conditions.
In particular, we are interested in the
transition probability density $p(x,t|x_0,t_0)$,
which is singled out by the initial condition (3.4),
because the asymptotic convergence in $L^1$ of
the solutions of equation (2.3) is ruled by
the asymptotic behaviour of $p(x,t|x_0,t_0)$
through the Chapman--Kolmogorov equation [1]
$$\rho (x,t)=\int_{-\infty}^{+\infty}p(x,t|y,t_0)
\rho (y,t_0^+)\,dy \, . \eqno{(4.8)}$$
It is clear at this point that
in every interval $[x_k,x_{k+1}]$ (both finite
or infinite) we can write the solution of
equation (2.3) along the guidelines
sketched in Sect. 3.
We must only keep
in mind that in $[x_k,x_{k+1}]$
we already know the invariant,
time--independent solution $\phi_n^2(x)$
which is never
zero inside the interval itself,
with the exception of the end
points $x_k$ and $x_{k+1}$.
Hence, as we have
seen in the general case, with the position
$$\rho (x,t)=\phi_n(x)g(x,t) \, , \eqno{(4.9)}$$
we can reduce (2.3) to the form
$$\partial_tg={\cal L}_ng
\, , \eqno{(4.10)}$$
where ${\cal L}_n$ is now the
self--adjoint operator defined on $[x_k,x_{k+1}]$ by
$${\cal L}_n\varphi(x)={{\rm d}\over{\rm d}x}
  \left[r(x){{\rm d}\varphi(x)\over{\rm d}x}\right]
  -q_n(x)\varphi(x) \, , \eqno{(4.11)}$$
with
$$r(x)=D>0\,;\qquad q_n(x)=
{v_{(+)n}^2(x)\over 4D}+{v_{(+)n}'(x)\over 2}\, .
\eqno{(4.12)}$$
Equation (4.10) is solved
by separating the
variables, so that we
immediately have
$\gamma(t)={\rm e}^{-\lambda t}$
while the spatial part $G(x)$ of $g$
must be solution of
$${\cal L}_nG(x) + \lambda
          G(x)=0 \, , \eqno{(4.13)}$$
with the boundary conditions
$$\bigl[2DG'(x)-v_{(+)n}(x)G(x)\bigr]_{x_k,x_{k+1}}
=0 \, . \eqno{(4.14)}$$
The general behaviour of the solutions of this
Sturm--Liouville problem obtained
as expansions in
the system of the eigenfunctions of (4.13) has already been
discussed in Section 3.
In particular we deduce from (3.12) that for
the stationary quantum states (more precisely, in every
subinterval defined by two subsequent nodes)
all the solutions of
(2.3) always converge in time
toward the correct quantum solution
$|\phi_n|^2$. As a further consequence, any quantum solution
$\phi^2_n$ defined on the entire interval
$(-\infty,+\infty)$ will
be stable under deviations from its initial condition.

\vskip 30pt
\noindent
{\bf 5. An explicit example: the harmonic oscillator}
\vskip 10pt
\noindent
To provide an explicit evolution
of the probability and
the transition probability
densities of stochastic mechanics,
we consider in detail the example
of a harmonic oscillator
associated to the potential
$$V(x)={m\over2}\,\omega^2x^2\,,\eqno{(5.1)}$$
with energy eigenvalues
$$E_n=\hbar\omega \left(n+{1\over2}\right) \, ,
\qquad n=0,1,2\dots \, .\eqno{(5.2)}$$
Introducing the notation
$$\sigma_0^2={\hbar\over2m\omega}\, , \eqno{(5.3)}$$
the time--independent part of the
eigenfunctions (4.1) reads
$$\phi_n(x)=
{1\over\sqrt{\sigma_0\sqrt{2\pi}2^nn!}} \,
{\rm e}^{-x^2/4\sigma_0^2}\,
H_n\left({x\over\sigma_0\sqrt{2}}\right)
\, , \eqno{(5.4)}$$
where $H_n$ are the Hermite polynomials.
The corresponding forward velocity fields
for the lowest lying levels are:
$$\eqalign{v_{(+)0}(x)&=-\omega x\,,\cr
v_{(+)1}(x)&=2\,{\omega\sigma_0^2\over x}-\omega x\,,\cr
v_{(+)2}(x)&=4\omega\sigma_0^2
\, {x\over x^2-\sigma_0^2}-\omega x\,,\cr} \eqno{(5.5)}$$
with singularities in the zeros
of the Hermite polynomials.
When $n=0$ the equation (2.3) takes the form
$$\partial_t\rho =\omega\sigma_0^2\partial_x^2 \rho
+\omega x\partial_x \rho +\omega \rho
\, , \eqno{(5.6)}$$
and the fundamental solution turns out to be
the Ornstein--Uhlenbeck
transition probability density
$$p_0(x,t|x_0,t_0)={1\over\sigma(t)\sqrt{2\pi}} \,
{\rm e}^{-[x-\alpha(t)]^2/2\sigma^2(t)} \, \, ,
\qquad(t\geq t_0)
\, , \eqno{(5.7)}$$
where we have introduced the notation
$$\alpha(t)=x_0\e^{-\omega(t-t_0)}\, \, , \qquad
\sigma^2(t)=\sigma_0^2\bigl[1-\e^{-2\omega(t-t_0)}\bigr]\,,
              \qquad(t\geq t_0)\,.\eqno{(5.8)}$$
The stationary Markov process associated to the
transition probability density (5.7)
is selected by the initial,
invariant probability density
$$\rho_0 (x)={1\over\sigma_0\sqrt{2\pi}}
\, \e^{-x^2/2\sigma_0^2} \, \, , \eqno{(5.9)}$$
which is also the asymptotic probability density
for every other initial condition when the
evolution is ruled by equation (5.6) (see [1])
so that the invariant distribution plays also
the role of the limit distribution.
Since this invariant probability density also
coincides with the quantum one
$\phi_0^2=|\psi_0|^2$,
the process associated by
stochastic mechanics to the ground state
of the harmonic oscillator
is nothing but the stationary
Ornstein--Uhlenbeck process.
\par
For $n\geq1$ the solutions of (2.3)
are determined in the following way.
As discussed
in the previous section, one has to solve the
eigenvalue problem (4.13) which
can now be written as
$$-\,{\hbar^2\over2m}\,G''(x)+
\left({m\over2}\omega^2x^2-\hbar\omega\,{2n+1\over2}
\right) G(x) = \hbar\lambda G(x) \, , \eqno{(5.10)}$$
in every interval $[x_k,x_{k+1}]$
between two subsequent singularities
of the forward velocity fields $v_{(+)n}$.
The boundary conditions at the end points of
these intervals, deduced
from (4.14) through (4.5), are
$$[\phi_nG'-\phi_n'G]_{x_k,x_{k+1}}
=0 \, . \eqno{(5.11)}$$
Reminding that $\phi_n$ (but not $\phi'_n$)
vanishes in $x_k,x_{k+1}$, the
conditions to impose are
$$G(x_k)=G(x_{k+1})=0 \, , \eqno{(5.12)}$$
where it is understood that for $x_0$
and $x_{n+1}$ we mean, respectively
$$\lim_{x\to-\infty}G(x)=0\, ,
\qquad\lim_{x\to+\infty}G(x)=0 \, .
\eqno{(5.13)}$$
It is also useful at this point
to state the eigenvalue problem in
adimensional form by using the reduced
eigenvalue
$\mu=\lambda/\omega$,
and the adimensional variable $x/\sigma_0$
which, by a slight abuse of notation, will be
still denoted by $x$.
In this way the equation (5.10) with the conditions
(5.12) becomes
$$\eqalign{y''(x)-\left({x^2\over4}
-{2n+1\over2}-\mu\right)y(x)&=0 \, , \cr
           y(x_k)=y(x_{k+1})&=0 \, . \cr}
\eqno{(5.14)}$$
If $\mu_m$ and
$y_m(x)$ are the eigenvalues and
eigenfunctions of (5.14),
the general solution of the corresponding
Fokker--Planck equation (2.3) will be
(reverting to dimensional variables)
$$\rho (x,t)=
\sum_{m=0}^{\infty}c_m\e^{-\mu_m\omega t}
\phi_n(x)y_m\left({x\over\sigma_0}\right)
\, . \eqno{(5.15)}$$
The values of the
coefficients $c_m$ will be fixed by the initial
conditions and by the obvious requirements
that $\rho(x,t)$ must be non negative
and normalized
during the whole time evolution.
Two linearly independent solutions of (5.14) are
$$y^{(1)}=\e^{-x^2/4}M
\left(-\,{\mu+n\over2},{1\over2};{x^2\over2}\right)
\, , \qquad y^{(2)}=
x\e^{-x^2/4}M \left(-\,{\mu+n-1\over2},
{3\over2};{x^2\over2}\right)\, ,
\eqno{(5.16)}$$
where $M(a,b;z)$ are the confluent
hypergeometric functions. The complete
specification of the solutions obviusly requires
the knowledge of all the eigenvalues $\mu_m$.
\par
We consider first the instance
$n=1$ ($x_0=-\infty$, $x_1=0$ and
$x_2=+\infty$), which can be completely
solved [5]. In this case
equation (5.14) must be
solved separately for $x\leq0$
and for $x\geq0$ with the boundary
conditions $y(0)=0$ and
$$\lim_{x\to-\infty}y(x)=
\lim_{x\to+\infty}y(x)=0 \, . \eqno{(5.17)}$$
A long calculation [5] shows that
the transition probability density
is now $$p(x,t|x_0,t_0)={x\over\alpha(t)}\,
{\e^{-[x-\alpha(t)]^2/2\sigma^2(t)}-
        \e^{-[x+\alpha(t)]^2/2\sigma^2(t)}
   \over \sigma(t)\sqrt{2\pi}} \, ,
\eqno{(5.18)}$$
where $\alpha(t)$ and $\sigma^2(t)$ are
defined in (5.8). It must be remarked
however that (5.18)
must be considered as restricted to $x\geq0$ when
$x_0>0$ and to $x\leq0$ when $x_0<0$,
and that only on these intervals
it is suitably normalized.
In order to take into account both
these possibilities we can
introduce the Heavyside function
$\Theta(x)$ so that for
every $x_0\not=0$ we will have
$$p_1(x,t|x_0,t_0)=\Theta(xx_0)\,{x\over\alpha(t)}\,
{\e^{-[x-\alpha(t)]^2/2\sigma^2(t)}-
        \e^{-[x+\alpha(t)]^2/2\sigma^2(t)}
 \over \sigma(t)\sqrt{2\pi}} \, . \eqno{(5.19)}$$
>From equation (4.8) we can
deduce the evolution
of every other initial probability
density. In particular it can be
shown that, with
$\rho_1(x)=\phi_1^2(x)$
$$\lim_{t\to+\infty}p_1(x,t|x_0,t_0)=2\Theta(xx_0)\,
 {x^2\e^{-x^2/2\sigma_0^2}\over\sigma_0^3\sqrt{2\pi}}=
   2\Theta(xx_0)\rho_1(x)\, . \eqno{(5.20)}$$
Hence, if $\rho (x,t_0^+)=\rho_{in} (x)$
is the initial
probability density, we have for $t>t_0$
$$\eqalign{\lim_{t\to+\infty}\rho (x,t)
& = \lim_{t\to+\infty} \int_{-\infty}^{+\infty}p(x,t|y,t_0)
\rho_{in} (y)\,dy\cr &=2\phi_1^2(x)\int_{-\infty}^{+\infty}
\Theta(xy)\rho_{in} (y)\,dy
=\Gamma(\epsilon ;x)\rho_1(x)\,,\cr}\eqno{(5.21)}$$
where we have defined the function
$$\Gamma(\epsilon  ;x) =
\epsilon  \Theta(x)+(2-\epsilon  )\Theta(-x)\,;\qquad
\epsilon  =2\int_0^{+\infty}\rho_{in} (y)\,dy \, .
\eqno{(5.22)}$$
When $\epsilon  =1$
(with symmetric initial probability, equally shared
on the two real semi--axis) we have $\Gamma(1;x)=1$
and the asymptotical probability density
coincides with the quantum
stationary density $\rho_1(x)=\phi_1^2(x)$.
If on the other hand
$\epsilon \not=1$ the asymptotic probability density
has the same shape of $\phi_1^2(x)$ but with
different weights on the two semiaxes.
\par
If we consider the higher excited states, the
Sturm--Liouville problem (5.14) must be solved
numerically in each sub--interval.
For instance, in the case
$n=2$ we have
$x_0=-\infty$, $x_1=-1$,
$x_2=1$
and
$x_3=+\infty$.
Considering in particular the
sub--interval
$[-1,1]$,
it can be shown that beyond $\mu_0=0$
the first few eigenvalues are determined
as the first possible values such that
$$M\left(-\,{\mu+1\over2} \, , \, {3\over2}
\, ; \, {1\over2} \, \right) = 0 \, .
             \eqno{(5.23)}$$
This gives
$\mu_1\sim 7.44$, $\mu_2\sim 37.06$, $\mu_3\sim 86.41$.

\vskip 30pt
\noindent
{\bf 6. Controlled evolutions}
\vskip 10pt
\noindent
In this Section we move on to
implement the program outlined in the introduction,
that is to exploit the transition probabilities of
Nelson stochastic mechanics to model
controlled quantum evolutions to arbitrarily assigned
final quantum states. We start by observing that
to every solution $\rho(x,t)$ of the
Fokker--Planck equation (3.1), with a
given $v_{(+)}(x,t)$ and constant diffusion coefficient (2.11),
we can always associate the wave function of a quantum system.
To this end, it is sufficient to
introduce a suitable time--dependent potential.
\par
Let us take a solution $\rho (x,t)$ of the Fokker--Planck
equation (3.1), with
a given $v_{(+)}(x,t)$ and a constant diffusion
coefficient $D$:
introduce the functions $R(x,t)$ and $W(x,t)$ from
$$\rho (x,t)=R^2(x,t)\,,\qquad\quad v_{(+)}(x,t)
=\partial_x W(x,t)\,,\eqno{(6.1)}$$
and remind from (2.12) that the relation
$$mv_{(+)}=\partial_x S+\hbar\,{\partial_x R\over R}=
\partial_x S+{\hbar\over2}\,{\partial_x \rho \over \rho }=
\partial_x\left(
S+{\hbar\over2}\ln\tilde \rho \right) \eqno{(6.2)}$$
must hold,
where $\tilde \rho $ is an adimensional
function (argument of a logarithm)
obtained from the probability density $\rho $ by means
of a suitable and arbitrary
dimensional multiplicative constant.
If we now impose
that the function $S(x,t)$ must be the phase of
a wave function as in (2.8), we immediately
obtain from (6.1) and (6.2) the equation
$$S(x,t)=mW(x,t)-{\hbar\over2}\ln
\tilde \rho (x,t)-\theta(t) \, , \eqno{(6.3)}$$
which allows to determine $S$ from $\rho$
and $v_{(+)}$ (namely $W$)
up to an additive arbitrary function
of time $\theta(t)$. However, in order that
the wave function (2.8) with $R$ and $S$
given above be a solution of a
Schr\"odinger problem in quantum mechanics,
we must also make sure that the Hamilton--Jacobi--Madelung
equation (2.9) is satisfied. Since $S$ and $R$ are
now fixed, equation (2.9) must be considered as a relation
(constraint) defining the controlling potential $V_{c}$,
which, after straightforward
calculations, yields
$$V_{c}(x,t)={\hbar^2\over4m}\,\partial_x^2\ln \tilde \rho +
    {\hbar\over2}\left(\partial_t\ln \tilde \rho
+ v_{(+)} \partial_x \ln \tilde \rho \right)
-{mv_{(+)}^2\over2}-m\partial_tW+\dot\theta \, .
\eqno{(6.4)}$$
Of course if we start with a quantum
wave function $\psi(x,t)$ associated to
a given time--independent potential
$V(x)$ and we select as a solution of (2.3)
exactly $\rho = |\psi|^{2}$, then the formula (6.4) always
yields back the given potential, as it should.
This can
be explicitly seen (to become familiar
with this kind of
approach) in the examples of the
ground state and the first excited state of the
harmonic oscillator potential (5.1),
by choosing respectively in equation (6.4)
$\theta(t)=\hbar\omega t/2$ and
$\theta(t)=3\hbar\omega t/2$, which
amounts to suitably fix the
zero of the potential energy.
\par
On the other hand the nonstationary fundamental
solution (5.7) associated to the
velocity field $v_{(+)0}(x)$ of (5.5) for
the case $n=0$
(we put $t_0=0$ to simplify
the notation) does not correspond to
a quantum wave function of the harmonic oscillator
whatsoever. However it is easy to show that, by choosing
$$\dot\theta(t)=
    {\hbar\omega\over2}
\left({2\sigma_0^2\over\sigma^2(t)}-1\right)=
 {\hbar\omega\over2}\,{1\over\tanh \omega t}\to
 {\hbar\omega\over2}\,,\qquad(t\to+\infty)\,,\eqno{(6.5)}$$
and the time--dependent controlling potential
$$V_{c}(x,t)=
{\hbar\omega\over2}
\left[{x-\alpha(t)\over\sigma(t)}\right]^2
{\sigma_0^2\over\sigma^2(t)}-{m\omega^2x^2\over2}\to
{m\omega^2x^2\over2}\, , \qquad(t\to+\infty)
\, , \eqno{(6.6)}$$
with $\alpha(t)$, $\sigma(t)$ and $\sigma_{0}$ defined
in equations (5.8) and (5.3),
we can define a quantum state, i.e.
a wave function $\psi_{c}(x,t)$
solution of the Schr\"odinger equation
in the potential (6.6).
At the same time $\psi_{c}$ is associated
to the transition probability density
of the form (5.7) which is its modulus squared.
Of course the fact that for $t\to+\infty$
we recover the harmonic potential
is connected to the already remarked fact that the usual quantum
probability density is also the limit distribution
for every initial condition and in particular also for
the evolution (5.7).
In the case $n=1$, with the $v_{(+)1}(x)$ as given by
equation(5.5)
and the transition probability density (5.19),
we define
$$T(x)={x\over\tanh x}\, , \eqno{(6.7)}$$
and then we choose
$$\dot\theta(t)={\hbar\omega\over2}
        \left({4\sigma_0^2\over\sigma^2(t)}-
   {2\sigma_0^2\alpha^2(t)\over\sigma^4(t)}-1\right)
\to{3\over2}\,\hbar\omega\,,\qquad(t\to+\infty)
\, , \eqno{(6.8)}$$
so that we have the following time--dependent
controlling potential (for every $x\not=0$):
$$\eqalign{V_{c}(x,t)&={m\omega^2x^2\over2}
      \left({2\sigma_0^4\over\sigma^4}-1\right)+
     \hbar\omega\left[1-{\sigma_0^2\over\sigma^2}\,
   T\left({x\alpha\over\sigma^2}\right)\right]-
{\hbar^2\over4mx^2}\, \left[1-T \left(
{x\alpha\over\sigma^2}\right)\right]\cr
&\to{m\omega^2x^2\over2}\, ,
\qquad\qquad(t\to+\infty)\, . \cr} \eqno{(6.9)}$$
The limit $t \to +\infty $ must
be obviously intended in a
physical sense, i.e. for times
much longer than $\lambda_1^{-1}$,
the largest characteristic time of
decay in the expansion (3.12). In
this particular case
$\lambda_1 = \omega $. In fact here too the
asymptotic potential is the usual one of
the harmonic oscillator, but it must
be considered separately on the positive and
on the negative $x$ semiaxis,
since in the point $x=0$ a singular behaviour would show up
when $t\to 0$.
This means that, also if asymptotically we recover
the right potential, it will be associated with a new boundary
condition in $x=0$ since the system must be
confined on one of the two semiaxes.

\vskip 30pt
\noindent
{\bf 7. Modelling transitions}
\vskip 10pt
\noindent
Given any couple
$(\rho, v_{(+)})$ associated
to a Fokker--Planck equation, the
possibility of promoting it to a
solution of a Schr\"odinger problem by a
suitable controlling
potential $V_{c}(x,t)$ enables to model
quantum evolutions driving,
for instance, the probability
density of a given quantum
stationary state to another (decays and
excitations). Moreover,
an immediate generalization of this scheme
might open the way to
modelling evolutions from a given, arbitrary
quantum state to an
eigenfunction of a given observable.
Besides other applications, this is something which
could become
a starting point for building simple models of the measurement
process, where one tries to dynamically describe the wave packet
collapse [10]. As a
first example let us consider the transition between the invariant
probability densities associated to the ground and the first
excited state of the harmonic oscillator potential (5.1):
$$\eqalign{\rho_0 (x)&=\phi_0^2(x) =
           {1\over\sigma_0\sqrt{2\pi}}
\, \e^{-x^2/2\sigma_0^2} \, , \cr
\rho_1 (x)&=\phi_1^2(x)=
           {x^2\over\sigma_0^3\sqrt{2\pi}}
\, \e^{-x^2/2\sigma_0^2} \, .
\cr}\eqno{(7.1)}$$
If we choose to describe the decay
$\phi_1 \to \phi_0 $ we may exploit
the Chapman--Kolmogorov equation (4.8) with the
transition probability density (5.7), and
the initial probability density
$\rho_{1}(x)$.
An elementary integration shows in this case
that the resulting evolution takes the form ($t_0=0$)
$$\rho_{1\to0} (x,t)=
\beta^2(t)\rho_0 (x)+\gamma^2(t)\rho_1 (x)
\, , \eqno{(7.2)}$$
where
$$\beta^2(t)=1-\e^{-2\omega t}\,,\qquad
    \gamma(t)=\e^{-\omega t} \, . \eqno{(7.3)}$$
Recalling $v_{(+)0} (x)$ as given in (5.5) and the evolving
probability density (7.2), and inserting them in equation (6.4) we
obtain the following form of the controlling potential:
$$V_{c}(x,t)={m\omega^2 x^2\over2}-2\hbar\omega
U(x/\sigma_0;\beta/\gamma) \, , \eqno{(7.4)}$$ where
$$U(x;b)={x^4+b^2x^2-b^2\over (b^2+x^2)^2} \, . \eqno{(7.5)}$$ The
parameter $$b^2(t)={\beta^2(t)\over\gamma^2(t)} = \e^{2\omega t}-1
\eqno{(7.6)}$$ is such that $b^2(0^+)=0$ and
$b^2(+\infty)=+\infty$. Thus $U$ goes to zero as $t\to+\infty$ for
any $x$, and as $t\to 0^+$ is 1 for every $x$, except for a
negative singularity in $x=0$. As a consequence, while for
$t\to+\infty$ the controlling potential (7.4) simply tends to the
potential (5.1), for $t\to0^+$ it presents an unessential shift of
$-2\hbar\omega$ in the zeroth level, and a deep negative
singularity in $x=0$.
\par
The singular behaviour of the controlling
potential at the initial
time of the evolution is a problem
connected to the proper
definition of the phase function $S$.
In fact, from (6.3) we have:
$$S(x,t)=
-{\hbar\over2}
\ln\left[\beta^2(t)+{x^2\over\sigma_0^2}
  \gamma^2(t)\right]
  -{\hbar\omega\over2}\, t \, ,\eqno{(7.7)}$$
so that in particular we have $$S(x,0^+)=-{\hbar\over2} \,
\ln{x^2\over\sigma_0^2} \, \, . \eqno{(7.8)}$$ We would instead
have expected that initially the phase be independent of $x$ as
for every stationary wave function. This means that in the
constructed evolution $S(x,t)$ presents a discontinuous behaviour
for $t\to 0^+$. The problem arises here from the fact that we
initially have a stationary state characterized by a probability
density $\rho_1 (x)$ and a velocity field $v_{(+)1}(x)$, and then
suddenly, in order to activate the decay, we impose to the same
$\rho_1$ to be embedded in a different velocity field $v_{(+)0}(x)$
which drags it toward a new stationary $\rho_0(x)$. This
discontinuous change from $v_{(+)1}$ to $v_{(+)0}$ is of course
responsible for the remarked discontinuous change in the phase of
the wave function. We have therefore modelled a transition which
starts with a sudden, discontinuous kick. To construct a
transition that evolves smoothly also for $t\to 0^+$ we should take
into account a continuous and smooth modification of the initial
velocity field into the final one. This requirement compels us to
consider a new class of Fokker--Planck equations with
time--dependent forward velocity fields $v_{(+)}(x,t)$. In
particular, to achieve the proposed smooth controlled decay
between two stationary states of the harmonic oscillator, we should
solve an evolution equation with a continuous velocity field
$v_{(+)}(x,t)$ which evolves smoothly from $v_{(+)1}(x)$ to
$v_{(+)0}(x)$. Clearly, the smoothing procedure can be realized in
several different ways and the selection must be dictated by the actual
physical requirements and outputs one is interested in.
A suitable smoothing for our transitions which leads to manageable
equations still has to be found; however in the
following Section we will study a problem in which the smoothness of
the evolution is {\it a priori} granted.

\vskip 30pt \noindent {\bf 8. Smooth transitions: coherent and
squeezed wave packets} \vskip 10pt \noindent As anticipated at the
end of the previous Section we will now consider an instance of
controlled evolution that does not require an extra smoothing procedure
for the driving velocity field, i.e. the transition between pairs
of coherent wave packets. In particular we will consider both the
transition from a coherent oscillating packet (coherent state) to
the ground state of the same harmonic oscillator, and a dynamical
procedure of squeezing a coherent wave packet.
\par
To this end we will recall a
simple result [1] which indicates how to find the
solutions of a particular class of evolution
equations (2.3) which includes the situation of
our proposed examples.
If the velocity field of the evolution
equation (2.3) has the linear form
$$v_{(+)}(x,t) = A(t) + B(t)x \eqno{(8.1)}$$
with $A(t)$ and $B(t)$ continuous functions of time,
then there are always solutions
of the Fokker-Planck equation in the normal form
${\cal N}\bigl(\mu(t), \nu(t) \bigr)$,
where $\mu(t)$ and $\nu(t)$ are solutions
of the differential equations
$$\dot\mu(t)-B(t)\mu(t)=A(t)\,;\qquad
     \dot\nu(t)-2B(t)\nu(t)=2D\eqno{(8.2)}$$
with suitable initial conditions.
The first case that we consider is
the coherent wave
packet with a certain initial displacement $a$:
$$\psi(x,t)=\biggl({1\over2\pi\sigma_0^2}\biggr)^{1/4}
\exp\biggl[-{(x-a\cos\omega t)^2\over4\sigma_0^2}
-i\biggl({4ax\sin\omega t-a^2\sin2\omega t
\over8\sigma_0^2}+{\omega t\over2}
\biggr)\biggr] \, , \eqno{(8.3)}$$
whose forward velocity field reads
$$v_{(+)}(x,t)=a\omega(\cos\omega t-\sin\omega t)
-\omega x\, . \eqno{(8.4)}$$
The field (8.4) is of the required form (8.1)
with $A(t)= a\omega(\cos\omega t-\sin\omega t)$
and $B(t)=-\omega$, while the configurational
probability density is
$$\rho (x,t)=|\psi(x,t)|^2
=\rho_0 (x-a\cos\omega t) \, , \eqno{(8.5)}$$
with $\rho_0$ is the one of the ground
state of the harmonic oscillator given by (7.1).
It is easy to show that when
$B(t)=-\omega$, as in the case of the
wave packets we are considering, there are
coherent solutions of (2.3) with
$\nu(t)=\sigma_0^2$ of the form ${\cal
N}\bigl(\mu(t),\sigma_0^2\bigr)$, i.e.
of the form $$\rho
(x,t)=\rho_0 \bigl(x-\mu(t)\bigr) \, .
\eqno{(8.6)}$$
Now the time evolution of such coherent
solutions can be determined in one
step, without implementing the
two--step procedure of first
calculating the transition probability
density and then, through
the Chapman-Kolmogorov equation,
the evolution of an arbitrary
initial probability density.
On the other hand if we compare (5.5)
and (8.4) we see that the
difference between $v_{(+)0}$ and
$v_{(+)}$
consists in the first,
time--dependent term of the
latter; hence it is natural
to consider the problem of solving
the evolution equation (2.3) with a
velocity field of the type
$$\eqalign{v_{(+)}(x,t)&=A(t)-\omega x \, , \cr
           A(t)&=a\omega(\cos\omega t
-\sin\omega t)F(t) \, , \cr} \eqno{(8.7)}$$
where $F(t)$ is an arbitrary function
varying smoothly between 1
and 0, or viceversa. In this case the evolution
equation (2.3)
still has coherent solutions of the form
(8.6) with a $\mu(t)$ dependent on
our choice of $F(t)$ through equation (8.2).
\par
A completely smooth transition from the
coherent, oscillating wave function
(8..3) to the ground state $\phi_0$
(5.4) of the harmonic oscillator
can now be achieved
for example by means of the
following choice of the function $F(t)$:
$$F(t)=1-\left(1-\e^{-\Omega t}\right)^N
=\sum_{k=1}^N(-1)^{k+1}{N\choose k}
\e^{-\omega_k t} \, \, , \eqno{(8.8)}$$
where
$$\Omega={\ln N\over \tau} \, ,
\qquad\omega_k=k\Omega\,;
\qquad \tau>0\,,\qquad N\geq 2 \, .
\eqno{(8.9)}$$
In fact, a function $F(t)$ of this form goes
monotonically from $F(0)=1$ to $F(+\infty)=0$
with a
flex point in $\tau$ (which can be considered as
the arbitrary instant of the transition)
where its derivative $F'(\tau)$ is negative and grows,
in absolute value, logarithmically with $N$.
The condition that the exponent $N\geq 2$
also guarantees that $F'(0)=0$,
and hence that the controlling
potential $V_{c}(x,t)$ given in equation (6.4)
will continuously start at
$t=0$ from the harmonic oscillator
potential (5.1), and asymptotically come back
to it for
$t\to+\infty$. Finally, the phase function $S(x,t)$
will too change continuously from
that of $\psi$ given in (8.3) to that of the
harmonic oscillator ground state $\psi_0$.
A long calculation yields
the explicit
form of the controlling potential:
$$V_{c}(x,t)=m\omega^2{x^2\over2}-m
\omega ax\sum_{k=1}^N(-1)^{k+1}{N\choose k}
\big[U_k(t)\omega_k\e^{-\omega_k t}
- W_k\omega\e^{-\omega t}\big] \, \, ,
\eqno{(8.10)}$$
where
$$\eqalign{U_k(t)&=\sin\omega
t+{2\omega^2\sin\omega t-\omega_k^2\cos\omega t
\over (\omega_k-\omega)^2+\omega^2}\,,\cr
 W_k&=1+{2\omega^2-
\omega_k^2\over(\omega_k-\omega)^2+\omega^2}
=\sqrt{2}U_k\left({\pi\over4\omega}\right) \, .
\cr} \eqno{(8.11)}$$
The parameters $\tau$ and $N$, apart
the constraints (8.9), are
free and can be fixed by the
particular form of the transition that
we want to implement, according to
what specific physical situations we are
interested in. Finally we remark that,
in a harmonic oscillator, the
transition between a coherent, oscillating wave
packet and the ground state is
a transition between a (Poisson)
superposition of all the energy
eigenstates to just one energy
eigenstate: an outcome which is
similar to that of an energy
measurement, but for the important
difference that here the result (the
energy eigenstate) is deterministically
controlled by a time--dependent potential.
The controlled transition that we have
constructed does not
produce mixtures, but
pure states (eigenstates) and may be considered
a dynamical model
for one of the branches of a
measurement leading to a selected
eigenvalue and eigenstate.
\par
Until now we have considered
transitions between gaussian wave packets
with constant width. However it is also
of great interest to discuss
the case of controlling potentials able
to produce a wave--packet evolution with
variable width: a kind of controlled squeezing of
the wave--packet.
This could be very useful in such instances as
the shaping of the Gaussian
output in the manufacturing of
molecular reactions, or
in the design of focusing
devices for beams in particle accelerators,
in which the width of the bunch must be properly
squeezed.
We will discuss now
a simple case which shows also that,
in the particular conditions chosen,
it is also possible to avoid the
integration of the differential equations (8.2).
\par
Let us remember that when the forward velocity
field has the form (8.1) the
Fokker-Planck equation (2.3) always
possesses Gaussian solutions of the form
$$\rho(x,t)={\e^{-[x-\mu(t)]^2/2\nu(t)}
\over \sqrt{2\pi\nu(t)}} \, , \eqno{(8.12)}$$
if $\mu(t)$ and $\nu(t)$  are solutions of (8.2).
We plan now to
describe evolutions of the quantum
state (2.8) such that I) both
$V_{c}(x,t)$ and $S(x,t)$ be
continuous and regular at every
instant, and II) the variance $\nu(t)$
satisfy the relations
$$\nu(-\infty)=\sigma^2_0
\, , \qquad\qquad\nu(+\infty)=\sigma^2_1\,.
\eqno{(8.13)}$$
In practice this means that,
if for example we require for the
sake of simplicity $\mu(t)=0$ at
every time, we will describe a
transition from the ground state
of an harmonic oscillator with
frequency $\omega_0=D/\sigma^2_0$
to the ground state of another
harmonic oscillator with
frequency $\omega_1=D/\sigma^2_1$. It is
convenient to remark here that this very
simple transition cannot
be achieved by means of an arbitrary
time--dependent potential
$V_{c}(x,t)$, given that
it goes from $m\omega_0^2x^2/2$ for
$t\to-\infty$
to  $m\omega_1^2x^2/2$ for $t\to+\infty$. The
intermediate evolution,
indeed, when not suitably designed, would
introduce components of every other
energy eigenstate of the final
harmonic oscillator which will not,
in general, asymptotically disappear.
\par
Let us recall here that the
relevant quantities are the phase function
$$S(x,t)=mW(x,t)-mD\ln
\tilde \rho (x,t)-\theta(t) \eqno{(8.14)}$$
(where $\theta(t)$ is arbitrary
and, from (6.1) and (8.1),
$W(x,t)=A(t)x+B(t)x^2/2$), and
the controlling potential
$$V_{c}(x,t)=mD^2\,\partial_x^2\ln
\tilde \rho + mD\left(\partial_t\ln
\tilde \rho + v_{(+)} \partial_x
\ln \tilde \rho \right)
-{mv_{(+)}^2\over2}-m\partial_tW+\dot\theta
\, . \eqno{(8.15)}$$
Both these two functions
are determined from the
knowledge of forward velocity field
$ v_{(+)}(x,t)$
and of the adimensional density
$\tilde \rho(x,t)=\sigma_0\rho(x,t)$. However
in this coherent evolution it will
not be necessary to integrate the
differential equations
(8.2) to obtain an explicit form of $S$ and $V_{c}$.
Indeed, since $A(t)$, $B(t)$
and $\ln\rho(x,t)$ can be
expressed through (8.2) in terms of $\mu(t)$, $\nu(t)$,
$\dot\mu(t)$, $\dot\nu(t)$ and $D$,
it is a straightforward matter to show that
the phase is of the general form
$$S(x,t)={m\over 2}\,\left[\Omega(t)x^2
-2U(t)x+\Delta(t)\right] \, , \eqno{(8.16)}$$
with
$$\eqalign{\Omega(t)&={\dot\nu(t)\over 2\nu(t)}
\, , \cr
        U(t)&={\mu(t)\dot\nu(t)-2\nu(t)\dot\mu(t)
\over 2\nu(t)}\,,\cr
           \Delta(t)&=D\,{\mu^2(t)\over\nu(t)}+D\,
\ln{2\pi\nu(t)\over\sigma_0^2}
    -{2\theta(t)\over m}\,,\cr}
     \eqno{(8.17)}$$
while the controlling potential reads $$V_{c}(x,t)={m\over 2}\,
\left[\omega^2(t)x^2-2a(t)x +c(t)\right] \, , \eqno{(8.18)}$$
where $$\eqalign{\omega^2(t)&={4D^2-2\nu(t)\ddot\nu(t)
+\dot\nu^2(t)\over 4\nu^2(t)}\,,\cr
  a(t)&={4D^2\mu(t)
+4\nu^2(t)\ddot\mu(t)-2\mu(t)\nu(t)\ddot\nu(t)
 +\mu(t)\dot\nu^2(t)\over 4\nu^2(t)}\,,\cr
   c(t)&={8D^2\mu^2(t)
-4D\nu(t)\dot\nu(t)-8D^2\nu(t)
 -\bigl(2\nu(t)\dot\mu(t)
-\mu(t)\dot\nu(t)+2D\mu(t)\bigr)\over
 4\nu^2(t)}+{2\dot\theta(t)\over m}\,.\cr}
   \eqno{(8.19)}$$
We can simplify our notation by
imposing that $\mu(t)=0$ (and
hence $\dot\mu(t)=\ddot\mu(t)=0$)
for every $t$, obtaining
$$S(x,t)={m\over
        2}\,\left[\Omega(t)x^2+\Delta(t)\right]
\, , \eqno{(8.20)}$$
with
$$\eqalign{\Omega(t)&={\dot\nu(t)\over 2\nu(t)}\,,\cr
           \Delta(t)&= D\,\ln{2\pi\nu(t)\over\sigma_0^2}
                          -{2\theta(t)\over m}\, . \cr}
          \eqno{(8.21)}$$
The controlling potential too reduces to $$V_{c}(x,t)={m\over 2}\,
\left[\omega^2(t)x^2 +c(t)\right] \, , \eqno{(8.22)}$$ where
$$\eqalign{\omega^2(t)&={4D^2-2\nu(t)\ddot\nu(t) +\dot\nu^2(t)
\over 4\nu^2(t)}\,,\cr c(t)&={2\dot\theta(t)\over m}
-D{\nu(t)\dot\nu(t)+2D\nu(t) \over\nu^2(t)}\,.\cr} \eqno{(8.23)}$$
Hence the evolution is completely defined, through the four
functions $\Omega(t)$, $\Delta(t)$, $\omega^2(t)$ and $c(t)$, by
$\theta(t)$ and $\nu(t)$. It is expedient in
particular to choose $$\theta(t)={mD\over2} \,
\ln{2\pi\nu(t)\over\sigma_0^2}+{mD^2t\over\nu(t)} \, .
\eqno{(8.24)}$$ In this way $$\Delta(t)= -{2D^2t\over\nu(t)} \, ,
\eqno{(8.25)}$$ because from (8.13) we have $\dot\nu(\pm\infty)=0$
so that (see (8.20)): $$\eqalign{S(x,t)&\sim
-{mD^2t\over\sigma_0^2}\,, \qquad\quad t\to-\infty \, , \cr
  S(x,t)&\sim -{mD^2t\over\sigma_1^2}\,,\qquad\quad
        t\to+\infty \, . \cr}\eqno{(8.26)}$$
This was to be expected from the fact that
$mD^2/\sigma_0^2=\hbar\omega_0/2$ and
$mD^2/\sigma_1^2=\hbar\omega_1/2$ are
the energy eigenvalues of
the ground states of the two harmonic
oscillators. Moreover from
the choice (8.24) also follows
that $c^2(\pm\infty)=0$ so that the
controlling potential (8.22)
will show no asymptotical extra terms with
respect to the initial and final harmonic
potentials.
\par
In order to completely specify the controlled
evolution we are now left with the determination
of the form of $\nu(t)$.
If $b=\sigma_1/\sigma_0$, then take
$$\nu(t)=\sigma_0^2\left({b+\e^{-t/\tau}
\over 1+\e^{-t/\tau}}\right)^2\,,
\qquad\quad(\tau>0) \, , \eqno{(8.27)}$$
so that the transition happens
around the instant $t=0$ and $\tau$
controls its velocity. We thus obtain the
explicit expressions
for the four functions (8.21)
and (8.23):
$$\eqalign{\Omega(t)&={b-1\over\tau}\,
{\e^{-t/\tau}\over(b+\e^{-t/\tau})(1+\e^{-t/\tau})}
\, , \cr \Delta(t)&=-{2D^2t\over\sigma_0^2}\,
              \left({1+\e^{-t/\tau}
\over b+\e^{-t/\tau}}\right)^2  \, , \cr
 \omega^2(t)&={D^2\over\sigma_0^4} \,
\left({1+\e^{-t/\tau}\over b+\e^{-t/\tau}}\right)^4+
{b-1\over\tau^2} \,
 {\e^{-t/\tau}(1-\e^{-t/\tau})
    \over(1+\e^{-t/\tau})^2(b+\e^{-t/\tau})}  \,,\cr
  c(t)&=-{4D^2(b-1)\over\sigma_0^2}\,
 {\e^{-t/\tau}(1+\e^{-t/\tau})\over(b+\e^{-t/\tau})^3}\,
              {t\over\tau}  \,.\cr}\eqno{(8.28)}$$
Their form is displayed in the Figures 1--4, where to fix an
example we have chosen the values $\tau=1$, $b=2$, $\omega_0=1$
and $\sigma_0=1$. As it can be seen in this case the behaviour of
the potential time--dependent parameters is not trivial even for
the very simple squeezing of a static gaussian wave packet from a
given variance to another. How to precisely follow this time
dependence in a stable way will be the argument of a forthcoming
paper, as discussed in the next section.

\vfill\eject \noindent {\bf 9. Conclusions and outlook} \vskip
10pt \noindent

We have shown how to treat the typical inverse problem in quantum
control, i.e. that of determining a controlling potential for a
given quantum evolution, in the framework of Nelson stochastic
mechanics. In this way we have been able to determine the general
characteristics of controlled evolutions between assigned initial
and final quantum states. The techniques of solution and the
relation between the transition probabilities, phase functions and
controlling potentials have been discussed on general grounds.
Detailed, explicit calculations have also been shown in the
paradigmatic test arena provided by the harmonic oscillator.
\par
Further extensions
of the method outlined in
the present paper are currently under
study. One immediate application to be faced
is the generalization of the analysis performed
for the harmonic oscillator to anharmonic
systems. The difficulty to be
faced on the way toward this aim
is that one is in general forced to
deal with approximate quantum wave functions,
as in the case of the quartic oscillator.
Therefore the controlled evolution must be
supplemented by a suitable feedback mechanism
ensuring that the error initially made in choosing
a certain initial approximate state does not
grow during the controlled time evolution.
One extremely interesting application would
be the description of a controlled evolution
driving initial approximate quantum states
of anharmonic systems to stable wave
packets generalizations of the coherent
states of the harmonic oscillator [15].
Besides the obvious interest in several
areas of quantum phenomenology, the above
is also of great potential interest in
discussing the control and the
reduction of aberrations in quantum--like
systems, i.e. deviations from the harmonic
evolution that are detected in systems like
charged beams in particle accelerators.
\par
Another very interesting future line of research that has been
left virtually unexplored in the present paper is the introduction
of optimization procedures. We have barely touched upon this
problem when discussing the smoothing of the controlled
transitions. Optimization of suitable functionals, chosen
according to the kind of physical evolution one needs or desires
to manufacture, would provide a powerful criterion of selection
among the different possible smoothed evolutions. Instances of
functionals to be optimized during the controlled dynamics that
come naturally to mind are the uncertainty products of conjugate
observables (to be optimized to a relative minimum under the
constraint of Schr\"odinger dynamics [16]), or the relative
entropy between the initial and final states. But many more can be
imagined and devised, according to the nature of the physical
problem considered.
\par
One last, but important consideration is in order.
When we implement a controlled
evolution by means of a suitable
controlling potential we must also bear in
mind that in practice small deviations away from the designed
potential and from the desired wave function are always possible.
In general such deviations are not subsequently reabsorbed but
rather tend to drag the state away from the required evolution.
Hence to really control these quantum evolutions it will be very
important to study their stability under small deviations
and perturbations: this is of crucial importance from
the standpoint of confronting the formal, theoretical scheme
with the practical applications.
Work is currently in progress in all the above mentioned extensions
of the present research, and we plan to soon report on it.

\vfill
\eject
\noindent
{\bf REFERENCES} \vskip 10pt
{\baselineskip 12pt
\item{[1]}  {Cufaro Petroni N and Guerra F 1995
                    {\it Found. Phys.} {\bf 25} 297}
\item{}     {Cufaro Petroni N 1995 Asymptotic behaviour of densities
                    for Nelson processes, in
                    {\it Quantum communications and measurement},
                    ed V P Belavkin et al (New York: Plenum Press)}
\item{}     {Cufaro Petroni N, De Martino S and De Siena S 1997 Non
                    equilibrium densities of Nelson
                    processes, in {\it New perspectives in the physics of
                    mesoscopic systems}, ed
                    S De Martino et al (Singapore: World Scientific)}

\item{[2]}  {Nelson E 1966 {\it Phys. Rev.} {\bf 150} 1079}
\item{}     {Nelson E 1967 {\it Dynamical Theories of Brownian Motion}
                    (Princeton: Princeton U.P.)}
\item{}     {Nelson E 1985 {\it Quantum Fluctuations} (Princeton:
                    Princeton U.P.)}
\item{}     {Guerra F 1981 {\it Phys. Rep.} {\bf 77} 263}
\item{}     {Blanchard Ph, Combe Ph and Zheng W 1987 {\it
             Mathematical and physical aspects of stochastic mechanics}
            (Berlin: Springer--Verlag)}

\item{[3]}  {Guerra F and Morato L M 1983
            {\it Phys. Rev.}  D {\bf 27} 1774}
\item{}     {Guerra F and Marra R 1983 {\it Phys. Rev.} D {\bf 28} 1916}
\item{}     {Guerra F and Marra R 1984 {\it Phys. Rev.}  D {\bf 29} 1647}
\item{}     {Morato L M 1985 {\it Phys. Rev.} D {\bf 31} 1982}
\item{}     {Loffredo M I and Morato L M 1989 {\it J. Math. Phys.}
            {\bf 30} 354}

\item{[4]}  {Bohm D and Vigier J P 1954 {\it Phys. Rev.} {\bf 96} 208}

\item{[5]}  {Cufaro Petroni N, De Martino S and
                    De Siena S 1998 {\it Phys. Lett.}  A {\bf 245} 1}

\item{[6]}  {Cufaro Petroni N 1989 {\it Phys. Lett.} A {\bf 141} 370}
\item{}     {Cufaro Petroni N 1991 {\it Phys. Lett.} A {\bf 160} 107}
\item{}     {Cufaro Petroni N and Vigier J P 1992 {\it Found. Phys.}
            {\bf 22} 1}

\item{[7]}  {Cufaro Petroni N, De Martino S, De Siena S
             and Illuminati F 1998 A stochastic model
             for the semiclassical collective dynamics of
             charged beams in partcle accelerators, in
            {\it 15th ICFA Advanced Beam Dynamics Workshop}
            (Monterey, California, US)}
\item{}     {Cufaro Petroni N, De Martino S, De Siena S, Fedele R,
             Illuminati F and Tzenov S 1998 Stochastic control
             of beam dynamics, in
            {\it EPAC'98 Conference} (Stockholm, Sweden)}

\item{[8]}  {Madelung E 1926 {\it Z.Physik} {\bf 40} 332}
\item{}     {de Broglie L 1926 {\it C.R.Acad.Sci.Paris} {\bf 183} 447}

\item{[9]}  {Bohm D 1952 {\it Phys.Rev.} {\bf 85} 166, 180}
\item{}     {Cufaro Petroni N, Dewdney C, Holland P,
             Kyprianidis T and Vigier J P 1985
            {\it Phys.Rev.} D {\bf 32} 1375}

\item{[10]} {Guerra F 1997 The problem of the physical
             interpretation of Nelson stochastic mechanics
             as a model for quantum mechanics, in
            {\it New perspectives in the physics of mesoscopic
                 systems} ed S De Martino et al
            (Singapore: World Scientific)}

\item{[11]} {Risken H 1989 {\it The Fokker-Planck equation}
            (Berlin: Springer)}

\item{[12]} {Tricomi F 1948 {\it Equazioni differenziali}
            (Torino: Einaudi)}

\item{[13]} {Tricomi F 1985 {\it Integral equations} (New York: Dover)}

\item{[14]} {Albeverio S and H\o gh-Krohn R 1974
            {\it J.Math.Phys.} {\bf 15} 1745}

\item{[15]} {De Martino S, De Siena S and Illuminati F 1997
            {\it J. Phys. A: Math. Gen.} {\bf 30} 4117}

\item{[16]} {Illuminati F and Viola L 1995
            {\it J. Phys. A: Math. Gen.} {\bf 28} 2953}
\par}

\vfill\eject \noindent
{\bf FIGURE CAPTIONS}\vskip 10pt
\item{Figure 1.}{The parameter $\Omega(t)$ in the phase function $S(x,t)$.}\vskip 10pt
\item{Figure 2.}{The parameter $\Delta(t)$ in the phase function $S(x,t)$.}\vskip 10pt
\item{Figure 3.}{The parameter $\omega^2(t)$ in the potential 
$V_{c}(x,t)$.}\vskip 10pt
\item{Figure 4.}{The parameter $c(t)$ in the potential 
$V_{c}(x,t)$.}\vskip 10pt

\bye